\input{aipcheck}

\documentclass[
    ,final            
  ]
  {aipproc}

\layoutstyle{8x11double}

\begin{document}

\title{VHE $\gamma$-ray Afterglow Emission from Nearby GRBs}


\author{P. H. Tam}{
  address={Landessternwarte (ZAH), K\"{o}nigstuhl 12, Heidelberg, D-69117, Germany}
}

\author{R. R. Xue}{
  address={Purple Mountain Observatory, Chinese Academy of
Sciences,
Nanjing 210008, China}
  ,altaddress={Graduate School, Chinese Academy of Sciences,
Beijing, 100012, China}
}

\author{S. J. Wagner}{
  address={Landessternwarte (ZAH), K\"{o}nigstuhl 12, Heidelberg, D-69117, Germany}
}

\author{B.~Behera}{
  address={Landessternwarte (ZAH), K\"{o}nigstuhl 12, Heidelberg, D-69117, Germany}
}

\author{Y.~Z.~Fan}{
  address={Purple Mountain Observatory, Chinese Academy of
Sciences, Nanjing 210008, China}
  ,altaddress={Niels Bohr International Academy, Niels Bohr Institute,
University of Copenhagen, Blegdamsvej 17, DK-2100 Copenhagen,
Denmark}
}

\author{D.~M.~Wei}{
  address={Purple Mountain Observatory, Chinese Academy of
Sciences, Nanjing 210008, China}
  ,altaddress={Joint Center for Particle Nuclear Physics
and Cosmology of Purple Mountain Observatory -- Nanjing University,
Nanjing 210008, China}
}

\begin{abstract}
Gamma-ray Bursts (GRBs) are among the potential extragalactic sources of
very-high-energy (VHE) $\gamma$-rays. We discuss the prospects of detecting VHE $\gamma$-rays with
current ground-based Cherenkov instruments during the afterglow phase. Using the fireball
model, we calculate the synchrotron self-Compton (SSC) emission from
forward-shock electrons. The modeled results are compared with
the observational afterglow data taken with and/or the sensitivity level of ground-based VHE
instruments (e.g. STACEE, H.E.S.S., MAGIC, VERITAS, and Whipple). We find that modeled SSC emission from bright and nearby bursts such as GRB~030329 are detectable by these instruments even with a delayed observation time of $\sim$10 hours.
\end{abstract}

\maketitle


\section{Introduction}
Gamma ray bursts (GRBs) are believed to be one of the cosmological
sources emitting GeV and higher energy photons. Tentative
evidence of distinct high-energy components has been accumulated
by EGRET's observations of several GRBs including GRB~940217 and GRB~941017~\cite{Hurley94,Gonzalez03}. In the fireball model, synchrotron emission of shock-accelerated
electrons is commonly considered to be responsible for prompt $\gamma$-ray emission as well as afterglow emission at lower energies, see, e.g. \cite{sari98}.
It is natural to expect that these photons are inverse-Compton up-scattered by electrons (e.g. synchrotron-self Compton, SSC), giving rise to a higher energy component peaking at GeV to TeV energies~\cite{wei98,sari01}.

Ground-based $\gamma$-ray instruments, such as
MAGIC, VERITAS, and H.E.S.S. are operating
at energies above $\sim$100 GeV. These Cherenkov instruments have very
large effective areas of $\sim 10^4-10^5~{\rm m^2}$ and a high rejection
rate of hadronic background, making them rather
sensitive to VHE $\gamma$-rays. Hence, these instruments are well suited to test the prediction of the above picture.

\section{The Model}
We employ the following leptonic jet model to calculate the afterglow emission from electrons accelerated in external forward shock, as described in~\cite{fan08}. The key elements of the model are summarized:
\begin{enumerate}
  \item a homogeneous
external medium with density $n$, or a wind profile $n\propto R^{-2}$;
  \item constant fractions of the shock energy given to
the electrons ($\epsilon_e$) and the magnetic field ($\epsilon_B$);
  \item electron distribution follows
$dN_\mathrm{e}/dE \propto E^{-p}$;
  \item energy injection in the form of
\begin{enumerate}
  \item $E_k \propto [1+(t/T)^2]^{-1}$; or
  \item $E_{k}\propto t^{1-q}$
\end{enumerate}
is used,
whenever necessary, to reproduce the flattening in
some observed afterglow light curves. Parameters $L_{\rm inj}$ (the rest-frame injected
luminosity) and injection timescale $T_\mathrm{inj}$ are also included;
  \item both synchrotron and inverse Compton emission are calculated.
\end{enumerate}
Klein-Nishina correction is required to estimate the VHE $\gamma$-ray emission. Attenuation by extra-galactic background light (EBL) is also included in the numerical calculations, using the `P0.45' EBL level considered in~\cite{aharonian06a}.


\begin{figure}
\includegraphics[bb= 0 0 300 218,clip,height=.3\textheight]{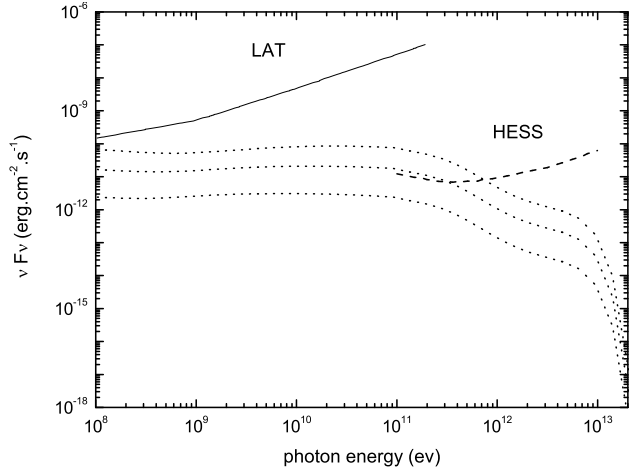}
\caption{Time-integrated SSC spectra from forward shocks,
integrated from 0.5 hour (top), 2 hours (middle) and 10 hours (bottom) after the
burst for 0.5 hour (dotted lines). Fermi/LAT~\citep{galli08} and H.E.S.S. (assuming photon spectral index $\Gamma=2.6$) sensitivities, both for an integration time of 0.5 hour, are also shown.}
\label{fig:example}
\end{figure}

\begin{table}
\begin{tabular}{lccc@{}ccc@{}cccc@{}c}
\hline               
GRB & $E_\mathrm{iso}$(erg) & $\theta_0$ & $n$(cm$^{-3}$)
& $p$ & $\epsilon_e$ & $\epsilon_B$ & $E_{\rm inj}$? & $L_{\rm inj}$ & $T_\mathrm{inj}$(s) & reference \\
\hline
 030329  &1.4$\times10^{53}$  &0.31 &100 &2.01 &0.1 & 0.001 & No & ... & ... & this work\\
 050509B &2.75$\times10^{48}$ &0.5  &1   &2.2  &0.15 &0.046 & No & ... & ... & \cite{bloom06}\\
 050709  &3.77$\times10^{50}$    &0.5  &6$\times10^{-3}$   &2.6  &0.4
 &0.25 & No & ... & ... & \cite{panaitescu06}\\
 051221A &$10^{52}$ &0.1  &0.01   &2.4  &0.3 &2$\times10^{-4}$ &
 Yes\tablenote{case (a)} &2$\times10^{48}$ &$<$1.5$\times$10$^4$ & \cite{fan06}\\
 060505  &2.6$\times10^{50}$  &0.4 &1 &2.1 &0.1 & 0.008 & No & ... & ... & \cite{xu09}\\
 060614  &5$\times10^{50}$  &0.08 &0.05 &2.5 &0.12 & 2$\times10^{-4}$ &
 Yes\tablenote{case (b) with $q=0$} &$10^{48}$ &$10^3$--2$\times$10$^4$ & this work\\
\hline
\end{tabular}
\caption{Model parameters for six nearby GRBs.}
\label{tab:parameters}
\end{table}

To understand the model prediction, first consider a factitious burst with the following parameter values: initial
(isotropic-equivalent) energy
$E_\mathrm{iso}=10^{51}$~erg,
initial half-opening angle $\theta_0= 0.4$, $n=1\,\rm cm^{-3}$, $p=2.2$,
 $\epsilon_{\rm e} =0.3$, $\epsilon_{\rm B} =0.01$, and
$z=0.16$. Time-integrated spectra (including both synchrotron and SSC components) from forward shocks are depicted in
 Figure~\ref{fig:example}. The modeled spectra are integrated from 0.5 hour, 2 hours, and
 10 hours after the burst for half an hour. One can see that for this factitious burst, current ground-based Cherenkov instruments such as H.E.S.S. would be more likely than \emph{Fermi}/LAT to probe the modeled emission.

\section{A sample of nearby GRBs}

We chose six nearby GRBs (i.e. $z<0.55$): GRB~030329, GRB~050509B,
GRB~050709, GRB~051221A, GRB~060505, and GRB~060614 in this study. The
available radio to X-ray afterglow
observation data were then used to obtain the parameter values in the model. Data from at least
two different wavebands were required. We
reproduced the multi-frequency afterglow behaviour of GRB~030329 and GRB~060614 using a set of parameters shown in
Table~\ref{tab:parameters} (c.f. \cite{Xue09} for details). For the other four GRBs, parameters were taken
from~\cite{bloom06}, \cite{panaitescu06}, \cite{fan06}, and \cite{xu09}.

\section{Comparison with observations}
\begin{figure}
\includegraphics[height=.28\textheight]{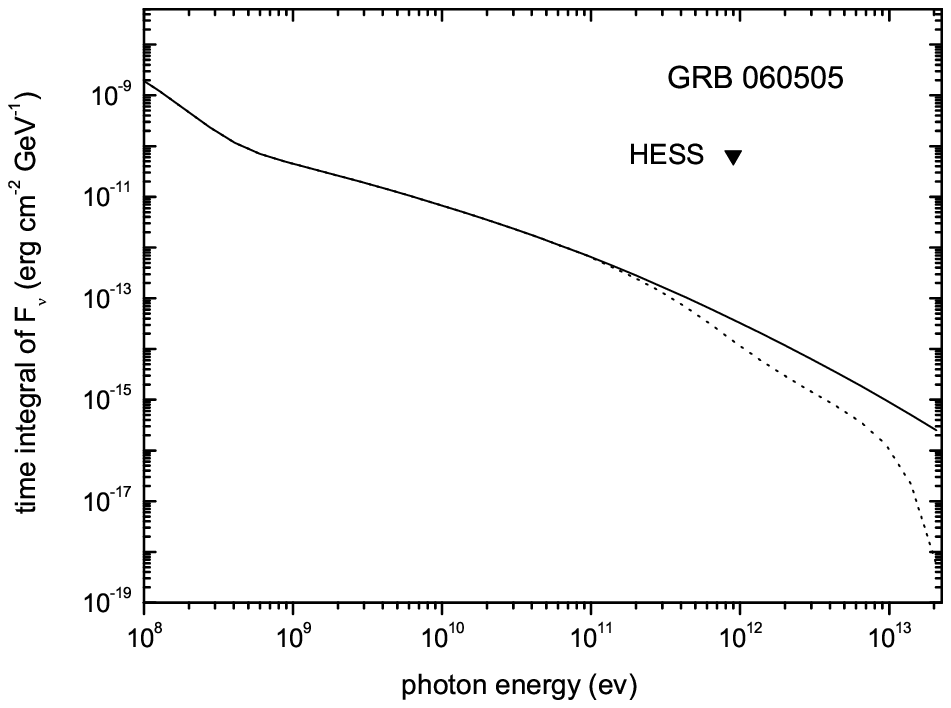}
\caption{Modeled time-integrated 0.1 GeV -- 20 TeV afterglow spectra of
GRB~030329, GRB~050509B and GRB~060505, in comparison with VHE upper
limits. Dotted and solid lines represent the spectra with and without
EBL-correction, respectively. The spectra were integrated over the
corresponding time intervals during which the upper limits were derived, as
shown in Table~\ref{tab:obs_time}. For GRB~030329, thick (upper) lines
indicate the modeled curve for the Whipple observation time, and thin
(lower) lines for the H.E.S.S. observation time. The upper limit points are
plotted at the corresponding average photon energies.}%
\label{fig:comparison}%
\end{figure}

Based on the parameters obtained as described in the previous section, the modeled GeV--TeV emission
was calculated. Figure~\ref{fig:comparison} shows the
modeled time-integrated 0.1 GeV -- 20 TeV afterglow spectra of GRB~030329,
GRB~050509B, and GRB~060505, where VHE $\gamma$-ray observational data (upper limits) are available for
comparison. The SSC spectra with and without EBL-correction are
shown for each GRB. The spectra were integrated over the respective time intervals during
which the upper limits were derived, as shown in Table~\ref{tab:obs_time}.

\begin{table}
\begin{tabular}{lcccrc}
\hline               
GRB      & $z$  & VHE instrument & $t_\mathrm{obs}-t_\mathrm{burst}$ & exposure & reference\\
\hline
 030329  &0.1685 & H.E.S.S.  & 11.5 days                       & 30 min   & \cite{tam08}\\
         &       & Whipple   & 64.55 hours                        & 65 min   & \cite{horan07}\\
 050509B &0.2248 & STACEE    & 20 min                          & 30 min   & \cite{jarvis08}\\
 060505  &0.089  & H.E.S.S.  & 19.4 hours                         & 120 min  & \cite{tam08}\\
\hline
\end{tabular}
\caption{Observational time intervals for three nearby GRBs}
\label{tab:obs_time}
\end{table}

\begin{figure}
\includegraphics[height=.28\textheight]{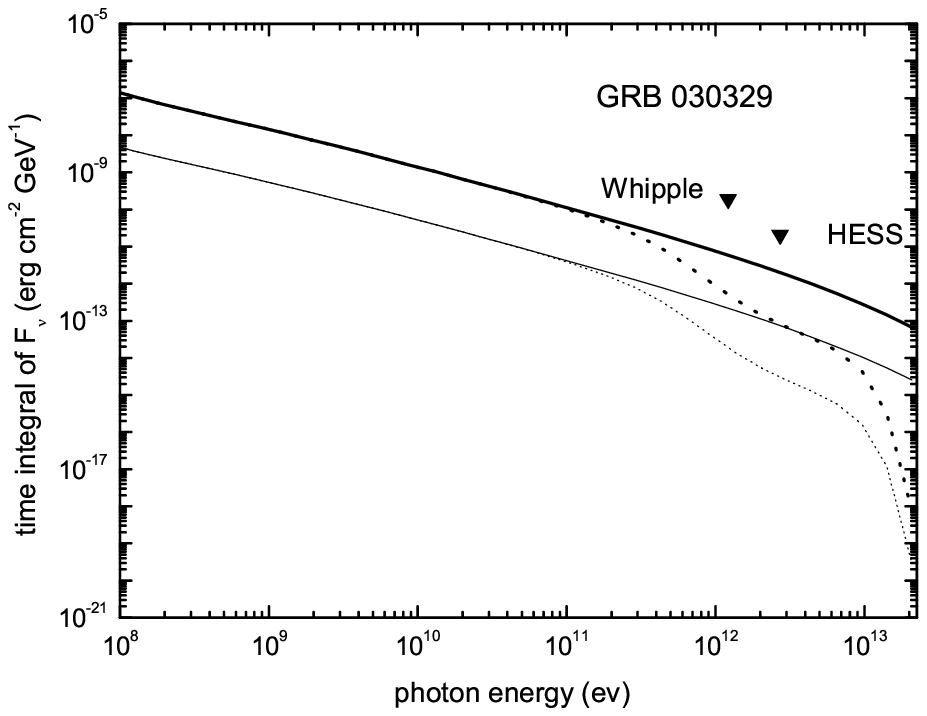}%
\includegraphics[height=.28\textheight]{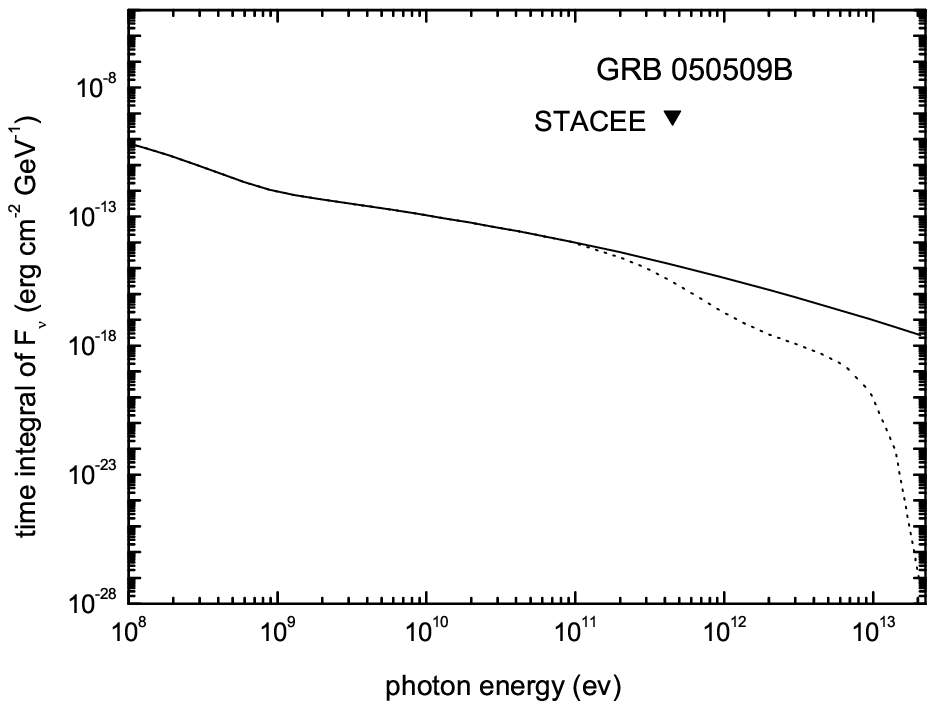}
\label{fig:comparison1}%
\end{figure}

\begin{figure}
\caption{Comparison between modeled VHE integral energy fluxes above 200~GeV
  (indicated by dots) for six nearby GRBs, and the H.E.S.S. sensitivity level of
  $6\times 10^{-12} \rm erg\,cm^{-2}\,s^{-1}$ (vertical line). The sensitivity flux level is defined as the minimum detectable point source flux for a 5 significance level detection in a 2-hour observation~\citep{aharonian06b}. The shaded region is above the sensitivity level. It is assumed that observations begin 10 hours after the burst.}
\includegraphics[height=.3\textheight]{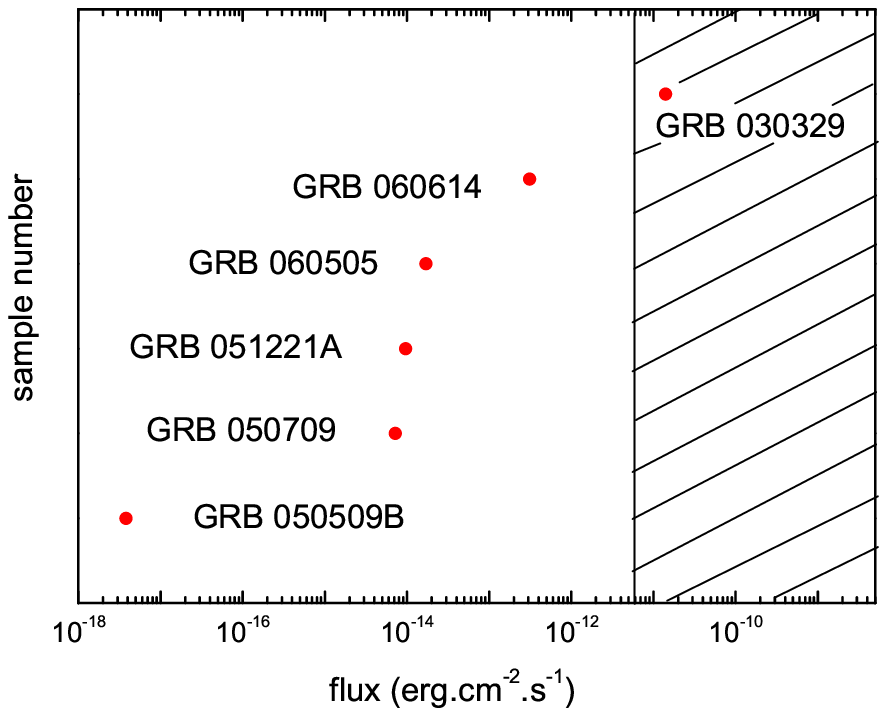}
\label{fig:10_hrs}
\end{figure}

Do we expect the current generation of VHE instruments, like MAGIC, H.E.S.S. or
VERITAS, to detect the predicted VHE $\gamma$-rays from nearby GRBs during the afterglow phase? Assuming that observations begin 10 hours after the burst and last for 2 hours, the EBL-attenuated energy fluxes above 200~GeV for these six GRBs, together with the H.E.S.S. sensitivity ($\Gamma$=2.6, same energy range), are shown in Figure~\ref{fig:10_hrs}.
\newpage
For GRB~030329, the expected energy flux is $1.4\times 10^{-11}$~erg~cm$^{-2}$~s$^{-1}$,
which is higher than the H.E.S.S. sensitivity level of $6\times 10^{-12}$ erg~cm$^{-2}$~s$^{-1}$~\cite{aharonian06b}. Therefore, predicted VHE $\gamma$-ray emission of
forward-shock electrons from a burst similar to GRB~030329 is detectable using
the current generation of VHE instruments.

\section{Conclusions}
In this proceeding, we discuss the prospects of detecting VHE $\gamma$-rays with
current ground-based instruments during the afterglow phase. The SSC emission model from
forward-shock electrons~\citep{fan08} was used to predict VHE $\gamma$-ray
emission in the afterglow phase. Klein-Nishina correction and EBL attenuation, both known to suppress the VHE $\gamma$-ray spectra, were taken into account. We chose a sample of six nearby
GRBs in this study. Our calculated results are
consistent with the upper limits obtained by the VHE $\gamma$-ray observations of
GRB~030329, GRB~050509B, and GRB~060505. Assuming observations taken 10
hours after the burst, the VHE $\gamma$-ray flux predicted from five GRBs is below the
sensitivity level of current Cherenkov instruments like MAGIC, H.E.S.S., and
VERITAS. However, for those bright and nearby bursts like GRB~030329, a VHE $\gamma$-ray detection is possible even with a delayed observation time of $\sim$10 hours after the burst.


\begin{theacknowledgments}
P.H. Tam acknowledges financial support from the \\
IMPRS `Astronomy and Cosmic
Physics at the University of Heidelberg'.
\end{theacknowledgments}



\bibliographystyle{aipproc}   



\end{document}